\documentclass[12pt]{article}%
\usepackage{graphicx}
\usepackage{amsmath}
\usepackage{amsfonts}
\usepackage{amssymb}%
\providecommand{\U}[1]{\protect\rule{.1in}{.1in}}

\begin{document}

\title{How the Kondo ground state avoids the orthogonality catastrophe}
\author{Gerd Bergmann\\Department of Physics\\University of Southern California\\Los Angeles, California 90089-0484\\e-mail: bergmann@usc.edu}
\date{\today}
\maketitle

\begin{abstract}
In the presence of a magnetic impurity the spin-up and down band states are
modified differently by the impurity. If the multi-electron scalar product
(MESP) between the occupied spin-up and down states approaches zero then this
defines an orthogonality catastrophe. In the present paper the MESP is
investigated for the FAIR (\textbf{F}riedel \textbf{A}rtificial \textbf{I}%
serted \textbf{R}esonance) solution for a Friedel-Anderson impurity. A basis
of Wilson states is used. The MESP is numerically determined for the
(enforced) magnetic, the singlet, and the triplet states as a function of the
number $N$ of Wilson states. The magnetic and the triplet state show an
exponentially decreasing MESP as a function of $N$. Surprisingly it is not the
number of states which causes this decrease. It is instead the energy
separation of the highest occupied state from the Fermi energy which
determines the reduction of the MESP. In the singlet state the ground-state
requires a finite MESP to optimize its energy. As a consequence there is no
orthogonality catastrophe. The MESP approaches a saturation value as function
of $N$.

PACS: 75.20.Hr, 71.23.An, 71.27.+a \newpage

\end{abstract}

\section{Introduction}

The orthogonality (or infrared) catastrophe was introduced and discussed
already 40 years ago \cite{A53}, \cite{H32}, \cite{Y6}, \cite{Y7}. An example
is a magnetic impurity in a metal host which interacts with the conduction
electron in the form $H^{\prime}=2J\left(  \mathbf{r}\right)  \mathbf{s\cdot
S}$. The effect of the z-component 2$J\left(  \mathbf{r}\right)  s_{z}S_{z}$
is the following. Let the spin direction of the impurity point upwards. Then
the wave function of the conduction electrons is pulled towards or pushed away
from the impurity, depending on the electron spin. As a consequence the scalar
product of corresponding s-electron states with opposite spin is slightly less
than 1.

If we denote the resulting (modified) bases for spin up and down as $\left\{
c_{\nu+}^{\dag}\right\}  $ and $\left\{  c_{\nu-}^{\dag}\right\}  $ with $N$
states in each basis ($1\leq\nu<N$) \ and if the lowest $N/2$ states are
occupied then the value of the multi-electron scalar product (\textbf{MESP})
between all occupied s-states with spin up and those with spin down is defined
by the determinant%
\[
M^{N/2}=\left\vert
\begin{array}
[c]{ccc}%
\left\langle c_{1,+}^{\dag}|c_{1,-}^{\dag}\right\rangle  &  & \left\langle
c_{1,+}^{\dag}|c_{N/2,-}^{\dag}\right\rangle \\
&  & \\
\left\langle c_{N/2,+}^{\dag}|c_{1,-}^{\dag}\right\rangle  &  & \left\langle
c_{N/2,+}^{\dag}|c_{N/2-}^{\dag}\right\rangle
\end{array}
\right\vert
\]

The common argument is that the multi-electron scalar product between all
occupied s-states with spin up and those with spin down approaches zero when
the number of occupied s-electron states approaches zero. If one compliments
this system of impurity with spin up plus polarized conduction electrons with
the time reversed system where all spin directions are reversed then a
transition between the two by spin-flip processes of the form $J\left(
\mathbf{r}\right)  \left[  s^{+}S^{-}+s^{-}S^{+}\right]  $ has vanishing
amplitude. Therefore the non-diagonal part of the $\mathbf{s\cdot S}$
interaction tries to prevent the orthogonality catastrophe. This can be well
traced in the Fair treatment of the Kondo impurity.

In the following the Friedel-Anderson (FA) impurity will be discussed where
this process is less obvious. The Hamiltonian for the FA-impurity is given by
\begin{equation}
H_{FA}=%
{\textstyle\sum_{\sigma}}
\left\{  \sum_{\nu=1}^{N}\varepsilon_{\nu}c_{\nu,\sigma}^{\dag}c_{\nu,\sigma
}+E_{d}d_{\sigma}^{\dag}d_{\sigma}+\sum_{\nu=1}^{N}V_{sd}(\nu)[d_{\sigma
}^{\dag}c_{\nu,\sigma}+c_{\nu,\sigma}^{\dag}d_{\sigma}]\right\}
+Un_{d\uparrow}n_{d\downarrow} \label{H_FA}%
\end{equation}

During the past few years the author has introduced a new numerical approach
to the Kondo and the FA-impurity problem, the FAIR-method (Friedel
Artificially Inserted Resonance) \cite{B151}, \cite{B152}\cite{B153}. It is
based on the fact the $n$-electron ground state of the Friedel Hamiltonian
(consisting of an electron band and a d-resonance) can be exactly expressed as
the sum of two Slater states \cite{B91}%
\begin{equation}
\Psi_{Fr}=Aa_{0}^{\dag}%
{\textstyle\prod\limits_{i=1}^{n-1}}
a_{i}^{\dag}\Phi_{0}+Bd^{\dag}%
{\textstyle\prod\limits_{i=1}^{n-1}}
a_{i}^{\dag}\Phi_{0}\label{psi_fr}%
\end{equation}
where $a_{0}^{\dag}$ is an artificial Friedel resonance state which determines
uniquely the full orthonormal basis $\left\{  a_{i}^{\dag}\right\}  $. An
extension of this ground state to the Friedel-Anderson and Kondo impurity
problem yields good numerical results. Recently this method was applied to
calculate the Kondo polarization cloud for those impurities \cite{B177}.

Three different solutions of the FA-Hamiltonian will be discussed: the
magnetic state, the singlet state and the triplet state. For sufficiently
large $U$ this Hamiltonian yields a magnetic state at temperature only above
the Kondo temperature $T_{K}$. However, a magnetic state can be enforced by
the structure of the variational state. This state will be called the
\textit{enforced magnetic state}. This avoids the finite temperature
treatment. This magnetic solution $\Psi_{MS}$ has the form
\begin{equation}
\Psi_{MS}=\left[  Aa_{0-\downarrow}^{\dag}a_{0+\uparrow}^{\dag}+Bd_{\downarrow
}^{\dag}a_{0+\uparrow}^{\dag}+Ca_{0-\downarrow}^{\dag}d_{\uparrow}^{\dag
}+Dd_{\downarrow}^{\dag}d_{\uparrow}^{\dag}\right]  \prod_{i=1}^{n-1}%
a_{i-\downarrow}^{\dag}\prod_{i=1}^{n-1}a_{i+\uparrow}^{\dag}\Phi
_{0}\label{PsiMS}%
\end{equation}
The coefficients $A,B,C,D$ and the compositions of the FAIR states
$a_{0+}^{\dag}$ and $a_{0-}^{\dag}$ are optimized to minimize the energy
expectation value of the FA Hamiltonian. Due to the condition $\left\langle
a_{i\tau}^{\dag}\Phi_{0}\left\vert H_{0}\right\vert a_{j\tau}^{\dag}\Phi
_{0}\right\rangle =0$ for $i,j>0$ and $\tau=+,-$ the FAIR states determine the
other states $a_{i\tau}^{\dag}$ of the basis $\left\{  a_{i\tau}^{\dag
}\right\}  $ uniquely.

The singlet state is a symmetric superposition of a magnetic state and its
time- (or spin-) reversed state, while the triplet state is the asymmetric
superposition. (The FAIR states $a_{0+}^{\dag}$ and $a_{0-}^{\dag}$ and the
coefficients $A,B,C,D$ are independently optimized for the magnetic, singlet,
and triplet states).

For the Fair solution the MESP between the occupied spin up and spin down
sub-bands is essentially given by
\begin{equation}
M_{+-}^{\left(  N/2\right)  }=\left\langle
{\textstyle\prod\limits_{i=0}^{N/2-1}}
a_{i+}^{\dag}\Phi_{0}\left\vert H_{0}\right\vert
{\textstyle\prod\limits_{j=0}^{N/2-1}}
a_{j-}^{\dag}\Phi_{0}\right\rangle \label{MESP}%
\end{equation}%
\[
\]

\section{Numerical Calculation of the Multi-Electron Scalar Product}

\subsection{ The enforced magnetic state}

For most of the numerical calculations Wilson states are used (see appendix
A). In the calculation the following parameters are used: $\left\vert
V_{sd}\right\vert ^{2}=0.1$, $U=1,E_{d}=-0.5$. The magnetic solution is
optimized for different numbers of Wilson states with $N=20,30,40,50,60.$
Table I shows $M_{MS}^{\left(  N/2\right)  }$ of the magnetic solution for the
different sizes $N$ of the bases. The third column gives the scalar product of
the two FAIR states, $\left\langle a_{0+}\Phi_{0}\mathbf{|}a_{0-}\Phi
_{0}\right\rangle _{MS},$ the fourth column the ground-state energy, and the
fifth column gives the magnetic moment. ($\Phi_{0}$ is the vacuum state). As
one can see the scalar product $\left\langle a_{0+}\Phi_{0}\mathbf{|}%
a_{0-}\Phi_{0}\right\rangle _{MS},$ the ground-state energy (in the enforced
magnetic state), and the moment have reached their final values already for
$N=30$. However, the multi-scalar product decreases with increasing $N$.%
\[%
\begin{tabular}
[c]{|l|l|l|l|l|l|}\hline
\textbf{N} & $M_{MS}^{\left(  N/2\right)  }$ & $\left\langle a_{0+}%
\mathbf{|}a_{0-}\right\rangle _{MS}$ & $E_{0,MS}$ & $\mu$ & $\left\langle
a_{N/2}\mathbf{|}a_{N/2}\right\rangle _{MS}$\\\hline
10 & 0.878 & 0.823 & -0.607799 & 0.514 & .92(.25)\\\hline
20 & 0.396 & 0.501 & -0.627446 & 0.687 & .64(.65)\\\hline
30 & 0.190 & 0.4852 & -0.62810 & 0.690 & .55(.69)\\\hline
40 & 0.0917 & 0.4845 & -0.62812 & 0.690 & .52(.76)\\\hline
50 & 0.0443 & 0.4845 & -0.62812 & 0.690 & .50(.77)\\\hline
60 & 0.0216 & 0.484 & -0.62812 & 0.690 & .49(.79)\\\hline
2*20 & 0.394 & 0.526 & -0.629323 & 0.66 & .64(.61)\\\hline
2*30 & 0.198 & 0.514 & -0.629779 & 0.67 & .57(.70)\\\hline
\end{tabular}
\ \ \
\]
$%
\begin{tabular}
[c]{l}%
Table I: The multi-electron scalar product (MESP) and other parameters for
the\\
Friedel-Anderson impurity in the enforced magnetic state. The different
columns\\
give the number of Wilson states, the MESP, the (single electron) scalar
product\\
$\left\langle a_{0+}\Phi_{0}\mathbf{|}a_{0-}\Phi_{0}\right\rangle _{MS}$
between the two FAIR states, the ground-state energy, and the\\
magnetic moment. The 6th column is explained in the text. The parameters
used\\
in the calculation are $\left\vert V_{sd}\right\vert ^{2}=0.1$, $U=1,E_{d}%
=-0.5$.
\end{tabular}
\ \ \ $%

\[
\]

In Fig.1 the logarithm of the multi-electron scalar product $\ln\left(
M_{MS}^{\left(  N/2\right)  }\right)  $is plotted versus the number of Wilson
states $N.$ It follows a straight line which corresponds to the relation%
\[
M_{MS}^{\left(  N/2\right)  }=1.\,\allowbreak7e^{-0.073\ast N}=1.\,\allowbreak
7\ast0.93^{-N}%
\]
Obviously, the multi-scalar product decreases exponentially with increasing
$N$.%

\[%
\begin{array}
[c]{cc}%
{\includegraphics[
height=2.455in,
width=3.1017in
]%
{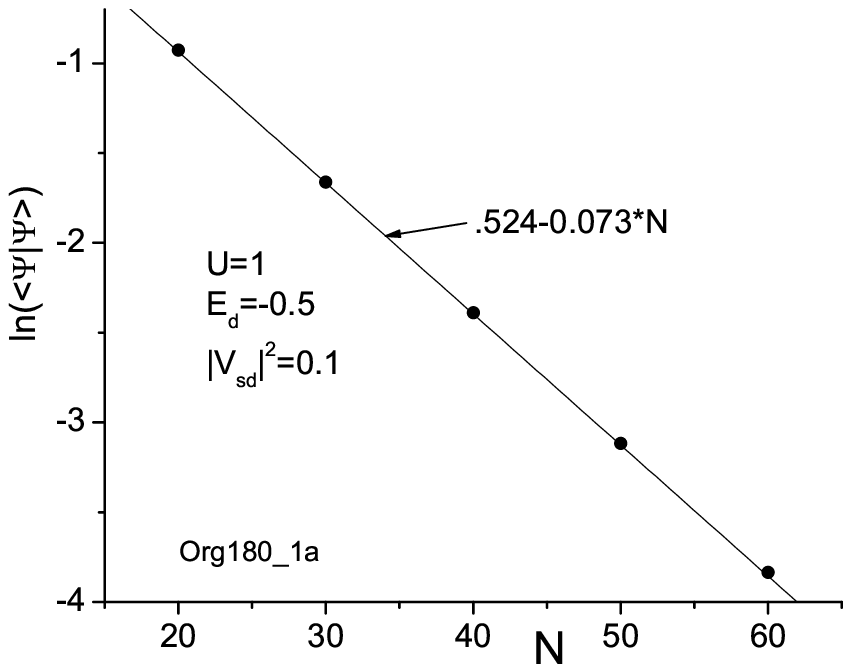}%
}%
&
\end{array}
\]

$\
\begin{tabular}
[c]{l}%
Fig.1: The logarithm of the multi-electron scalar product MESP\\
$\left\langle \prod_{i=0}^{n-1}a_{i+}^{\dag}\Phi_{0}|\prod_{i=0}^{n-1}%
a_{i-}^{\dag}\Phi_{0}\right\rangle $ is plotted versus the number of Wilson\\
states $N$ with $n=N/2$ for the magnetic state.
\end{tabular}
\ $%
\[
\]

In the next step I check whether it is just the number of states $N$ which
reduces $M_{MS}^{\left(  N/2\right)  }$. For this purpose the $N$ energy cells
for $N=20$ and $30$ are sub-divided into two. This is achieved by using
$\Lambda=\sqrt{2}.$ This doubles the number of Wilson states but adds only one
state (for positive and negative energy) closer to the Fermi level. The
results of this calculation are collected at $N=2\ast20$ and $2\ast30$. It
turns out that the doubling has essentially only a minor effect on
$M_{MS}^{\left(  N/2\right)  }$. This is on a first sight rather surprising
since it was believed that the increase of the number of states causes the
orthogonality catastrophe of the MESP.

To further confirm this observation I take the energy frame with $N=20$ and
subdivide the energy range $\left(  -1:-1/4\right)  $ into cells with a width
of 1/8, replacing two Wilson states by six new states. (The same is done for
the positive range). This changes $M_{MS}^{N/2}$ from $0.396$ to $0.405$.
Splitting the same energy range into 14 cells with a width of $1/32$ yields
the MESP $M_{MS}^{N/2}=0.411$. This shows that increasing $N$ by subdividing
an energy range does not contribute to an orthogonality catastrophe (as long
as the energy range does not border the Fermi level at the energy $0$).

On the other hand, the smallest (absolute) energies have a great impact on the
MESP. To investigate this question further I take the energies for $N=20$ and
shift the two states which are closest to the Fermi level towards the Fermi
level. The four energy cells which are closest to the Fermi level are
$\mathfrak{C}_{9}=\left(  -2^{-8}:-2^{-9}\right)  $, $\mathfrak{C}%
_{10}=\left(  -2^{-9}:0\right)  $, $\mathfrak{C}_{11}=\left(  0:2^{-9}\right)
$, $\mathfrak{C}_{12}=\left(  2^{-9}:2^{-8}\right)  $. I replace $\pm2^{-9}$
by $\pm2^{-19}$. Then the (average) energies of the corresponding states are
$\varepsilon_{9}=-\frac{2049}{1048\,576}\thickapprox$ $-1.\,\allowbreak
954\,1\times10^{-3},$ $\varepsilon_{10}=-2^{-20},$ $\varepsilon_{11}=2^{-20}$
and $\varepsilon_{12}=1.\,\allowbreak954\,1\times10^{-3}$. Of course this
reduces the s-d interaction strength $V_{sd}\left(  \nu\right)  $ for
$\nu=10,11$ from $\left[  2^{-9}/2\right]  ^{1/2}=2^{-5}$ to $\left[
2^{-19}/2\right]  ^{1/2}=2^{-10}$. (The ratio $\left\vert V_{sd}\left(
\nu\right)  \right\vert ^{2}$ /$\varepsilon_{\nu}$ remains constant).

After optimizing the $\left\{  a_{i+}\right\}  $ and $\left\{  a_{i-}\right\}
$ bases and the coefficients $A,B,C,D$ the resulting MESP is reduced to
$M_{MS}^{10}=0.0208$. The number of states is still $N=20$. The shifting of
the smallest energies from $\pm2^{-10}$ to $\pm2^{-20}$ changes the value of
the MESP from 0.396 to 0.0208. This shows that the value of the MESP is
determined by the occupied state closest to the Fermi level. The total number
of states is only important when it determines the energy of this state.%

\[
\]

\subsection{The singlet state}

In the next step I calculate the MESP for the singlet ground state. The same
parameters $\left\vert V_{sd}\right\vert ^{2}=0.1$, $U=1,E_{d}=-0.5$ are used
as in table I and Fig.1. The FAIR solution for the singlet state is obtained
by reversing all spins in $\Psi_{MS}$ and combining the two states.%
\[
\Psi_{SS}=\Psi_{MS}\left(  \uparrow\downarrow\right)  +\Psi_{MS}\left(
\downarrow\uparrow\right)
\]%
\begin{align}
&  =\left[  Aa_{0-\downarrow}^{\dag}a_{0+\uparrow}^{\dag}+Bd_{\downarrow
}^{\dag}a_{0+\uparrow}^{\dag}+Ca_{0-\downarrow}^{\dag}d_{\uparrow}^{\dag
}+Dd_{\downarrow}^{\dag}d_{\uparrow}^{\dag}\right]  \prod_{i=1}^{n-1}%
a_{i-\downarrow}^{\dag}\prod_{i=1}^{n-1}a_{i+\uparrow}^{\dag}\Phi
_{0}\label{PsiSS}\\
&  +\left[  A^{\prime}a_{0-\uparrow}^{\dag}a_{0+\downarrow}^{\dag}+B^{\prime
}d_{\uparrow}^{\dag}a_{0+\downarrow}^{\dag}+C^{\prime}a_{0-\uparrow}^{\dag
}d_{\downarrow}^{\dag}+D^{\prime}d_{\uparrow}^{\dag}d_{\downarrow}^{\dag
}\right]  \prod_{i=1}^{n-1}a_{i-\uparrow}^{\dag}\prod_{i=1}^{n-1}%
a_{i+\downarrow}^{\dag}\Phi_{0}\nonumber
\end{align}

The coefficients $A,B,C,D,A^{\prime},B^{\prime},C^{\prime},D^{\prime}$ and the
compositions of the FAIR states $a_{0+}^{\dag}$ and $a_{0-}^{\dag}$ are again
optimized to minimize the energy expectation value of the FA Hamiltonian. In
table II are the corresponding data collected. Again the first four columns
give the same data as in table I, i.e. the number of Wilson states, the MESP,
the (single electron) scalar product between the two FAIR states $a_{0+}%
^{\dag}$ and $a_{0-}^{\dag}$ and the ground-state energy. The 5th column gives
the Kondo energy (difference between the relaxed triplet energy and the
singlet ground-state energy).

\textbf{Dependence on the number of states }$\mathbf{N}$%

\[%
\begin{tabular}
[c]{|l|l|l|l|l|l|}\hline
\textbf{N} & $M_{SS}^{\left(  N/2\right)  }$ & $\left\langle a_{0+}%
\mathbf{|}a_{0-}\right\rangle _{SS}$ & $E_{0,SS}$ & $\Delta E$ &
$>$%
.999\\\hline
10 & 0.749 & 0.645 & -0.62272 & $1\allowbreak4.8\times10^{-3}$ & \\\hline
20 & 0.742 & 0.6448 & -0.637535 & $1\allowbreak0.1\times10^{-3}$ &
9-10\\\hline
30 & 0.742 & 0.6448 & -0.637965 & $9.\,\allowbreak87\times10^{-3}$ &
9-21\\\hline
40 & 0.742 & 0.6448 & -0.63798 & $9.\,\allowbreak86\times10^{-3}$ &
9-31\\\hline
50 & 0.742 & 0.6448 & -0.63798 & $9.\,\allowbreak86\times10^{-3}$ &
9-41\\\hline
60 & 0.742 & 0.6448 & -0.637973 & $9.\,\allowbreak85\times10^{-3}$ &
9-51\\\hline
2*20 & 0.751 & 0.657 & -0.639684 & $1\allowbreak0.4\times10^{-3}$ &
17-23\\\hline
2*30 & 0.751 & 0.657 & -0.639993 & $1\allowbreak0.2\times10^{-3}$ &
18-43\\\hline
\end{tabular}
\]

$%
\begin{tabular}
[c]{l}%
Table II: The multi-electron scalar product (MESP) and other parameters for\\
the Friedel-Anderson impurity in the singlet state. The different columns
give\\
the number of Wilson states $N$, the MESP, the (single electron) scalar
product\\
$\left\langle a_{0+}\Phi_{0}\mathbf{|}a_{0-}\Phi_{0}\right\rangle _{MS}$
between the two FAIR states, the ground-state energy, and the\\
Kondo energy. The 6th column is explained in the text. The parameters used\\
in the calculation are $\left\vert V_{sd}\right\vert ^{2}=0.1$, $U=1,E_{d}%
=-0.5$.
\end{tabular}
\ \ \ $%
\[
\]

For the last column I calculated the scalar product $\left\langle
a_{+,i}^{\dag}\Phi_{0}|a_{-,j}^{\dag}\Phi_{0}\right\rangle $ for all pairs of
$\left(  i,j\right)  $ which form a $N\times N$-matrix. It turns out that the
diagonal elements $\left\langle a_{+,i}^{\dag}\Phi_{0}|a_{-,i}^{\dag}\Phi
_{0}\right\rangle $ close to the Fermi energy approach the value one. For
example\ if the sixth column shows for $N=30$ the value $"9-21"$ then the
values of the (single particle) scalar products $\left\langle a_{+,i}^{\dag
}\Phi_{0}|a_{-,i}^{\dag}\Phi_{0}\right\rangle $ lie between 0.999 and 1.000
for $9\leq i$ $\leq21$. Obviously the states $a_{+,i}^{\dag}$ and
$a_{-,i}^{\dag}$ are almost identical in this interval. This is very different
for the enforced magnetic state. There, in table I the 6th column shows the
value of the diagonal scalar product for $i=N/2$ and the larger one of its two
neighbors $\left\langle a_{+,N/2}^{\dag}\Phi_{0}|a_{-,N/2\pm1}^{\dag}\Phi
_{0}\right\rangle .$

\textbf{Dependence on the interaction }$\left\vert V_{sd}\right\vert ^{2}$

The MESP in the singlet state depends on the strength of the s-d interaction.
Keeping the number of Wilson states constant $N=40$ the MESP is numerically
determined and collected in table III.
\[%
\begin{tabular}
[c]{|l|l|l|l|l|l|}\hline
$\left\vert \mathbf{V}_{sd}\right\vert ^{2}$ & $M_{SS}^{\left(  N/2\right)  }$
& $\left\langle a_{0+}^{\dag}\mathbf{|}a_{0-}^{\dag}\right\rangle _{SS}$ &
$E_{0,SS}$ & $\Delta E$ &
$>$%
.999\\\hline
0.10 & 0.742 & 0.645 & -0.637977 & $9.\,\allowbreak86\times10^{-3}$ &
-\\\hline
0.09 & 0.708 & 0.604 & -0.621392 & $8.\,\allowbreak13\times10^{-3}$ &
10-30\\\hline
0.08 & 0.665 & 0.553 & -0.605078 & $6.\,\allowbreak14\times10^{-3}$ &
10-30\\\hline
0.07 & 0.607 & 0.489 & -0.589200 & $4.\,\allowbreak09\times10^{-3}$ &
11-29\\\hline
0.06 & 0.527 & 0.406 & -0.573975 & $2.\,\allowbreak22\times10^{-3}$ &
11-29\\\hline
0.05 & 0.407 & 0.299 & -0.559655 & $\allowbreak8.\,\allowbreak21\times10^{-4}$
& 12-28\\\hline
\end{tabular}
\ \
\]
$%
\begin{tabular}
[c]{l}%
Table III: The multi-electron scalar product (MESP) and other parameters for\\
the Friedel-Anderson impurity in the singlet state. The columns give the\\
s-d interaction $\left\vert V_{sd}\right\vert ^{2}$, the MESP, the scalar
product $\left\langle a_{0+}^{\dag}\Phi_{0}|a_{0-}^{\dag}\phi_{0}\right\rangle
$, the\\
ground-state energy, and Kondo energy $\Delta E.$ The 6th column is explained
in\\
the text. The parameters used in the calculation are $U=1,E_{d}=-0.5$ and\\
the number of Wilson states is $N=40$.
\end{tabular}
\ \ $%
\[
\]

In Fig.2 the logarithm of the Kondo energy is plotted versus the logarithm of
the MESP. A linear dependence is obtained. The MESP shows a weak dependence on
the Kondo energy with a power of about $1/4$.%

\[%
\begin{array}
[c]{cc}%
{\includegraphics[
height=2.4989in,
width=3.0353in
]%
{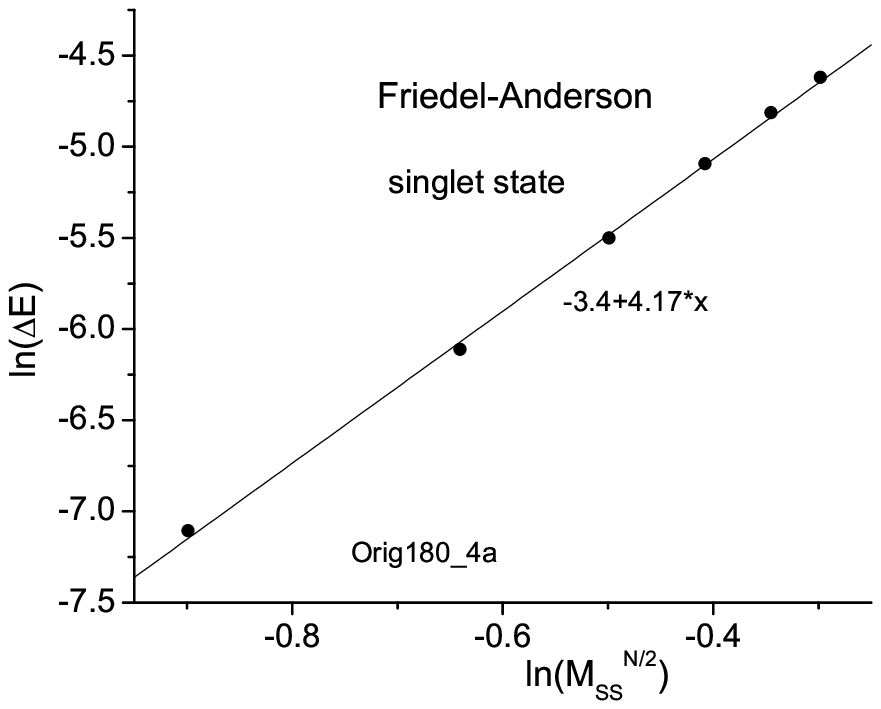}%
}%
&
\end{array}%
\begin{tabular}
[c]{l}%
Fig.2: The log-log plot of the\\
Kondo energy versus the MESP\\
for the singlet state for different\\
s-d interactions.
\end{tabular}
\
\]%
\[
\]

\subsection{The triplet state}

If one arranges the spins in each component in equ. (\ref{PsiSS}) for the
singlet state so that all spin-down creation operators are moved to the left
and all spin-up ones to the right, then the coefficients are pair-wise equal,
i.e. $A^{\prime}=A$, $B^{\prime}=B,$etc. This yields the symmetric or singlet
state. On the other hand, if one sets the coefficients pair-wise opposite
equal, i.e. $A^{\prime}=-A$, $B^{\prime}=-B,$etc then one obtains the
asymmetric or triplet state. Of course, one has to restart the optimization of
$a_{0,+}^{\dag}$, $a_{0,-}^{\dag}$ and $A,B,C,D$. For the asymmetric state a
finite MESP increases the total energy. So if one searches for the relaxed
triplet state with minimal energy one may expect a strong reduction of the
MESP. This is indeed found. In table III the data for the triplet state are
collected for the same parameters as before.%

\[%
\begin{tabular}
[c]{|l|l|l|l|l|}\hline
\textbf{N} & $M_{TS}^{\left(  N/2\right)  }$ & $\left\langle a_{0+}%
\mathbf{|}a_{0-}\right\rangle _{TS}$ & $E_{0,SS}$ & $\left\langle
a_{N/2}\mathbf{|}a_{N/2}\right\rangle _{TS}$\\\hline
10 & 0.926 & 0.891 & -0.582822 & .97(.12)\\\hline
20 & 0.0998 & 0.448 & -0.626428 & .16(.85)\\\hline
30 & 0.0162 & 0.481 & -0.628054 & .049(.88)\\\hline
40 & 2.59$\times10^{-3}$ & 0.484 & -0.628117 & .020(.89)\\\hline
50 & $3.02\times10^{-4}$ & 0.484 & -0.628119 & .082(.96)\\\hline
60 & 4.17$\times10^{-5}$ & 0.484 & -0.628119 & .21(.95)\\\hline
\end{tabular}
\
\]

$%
\begin{tabular}
[c]{l}%
Table III: The multi-electron scalar product (MESP) and other parameters for\\
the Friedel-Anderson impurity in the triplet state. The columns give the
number\\
of Wilson states $N$, the MESP, $\left\langle a_{0,+}^{\dag}\Phi_{0}%
|a_{0,-}^{\dag}\Phi_{0}\right\rangle $, and the ground-state energy.\\
The last column is explained in the text. The parameters used in the
calculation\\
are $\left\vert V_{sd}\right\vert ^{2}=0.1$, $U=1,E_{d}=-0.5$.
\end{tabular}
\ \ $%
\[
\]

In the limit of large $N$ the MESP approaches zero. That means that the two
components in equ. (\ref{PsiSS}) (top and the bottom line) are completely
decoupled. The triplet state consists of two orientations of the magnetic
state with zero interaction between them. Therefore the energies in the
magnetic and the triplet state become equal for large $N$. A comparison
between table I and table III does indeed show this agreement.

The dependence of the logarithm of the MESP on the number $N$ of Wilson states
is plotted in Fig.3.%
\[%
\begin{array}
[c]{cc}%
{\includegraphics[
height=2.6808in,
width=3.2569in
]%
{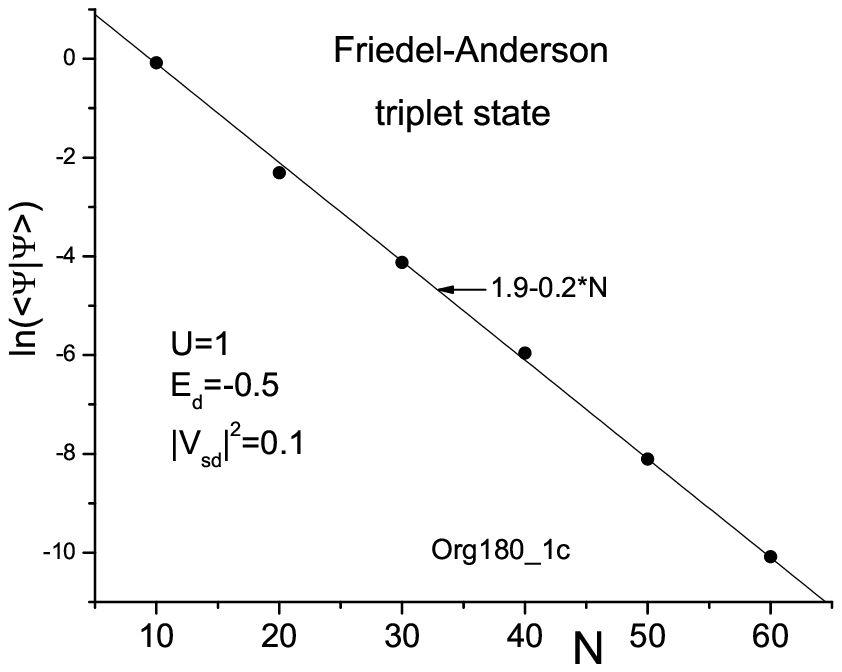}%
}%
&
\end{array}%
\begin{tabular}
[c]{l}%
Fig.3: The logarithm of the MESP\\
$\left\langle \prod_{i=0}^{n-1}a_{i+}^{\dag}\Phi_{0}|\prod_{i=0}^{n-1}%
a_{i-}^{\dag}\Phi_{0}\right\rangle $ is plotted\\
versus the number of Wilson states $N$\\
with $n=N/2$ for the triplet state
\end{tabular}
\]%
\[
\]

\section{Discussion and Conclusion}

The states $a_{+,i}^{\dag}$ are constructed from the basis $c_{\nu}^{\dag}$ by
extracting a FAIR state $a_{+,0}^{\dag}$. Therefore the states $a_{+,i}^{\dag
}$ and $c_{\nu}^{\dag}$ are pair-wise quite similar except that there is one
state missing in the basis $\left\{  a_{+,i}\right\}  $.
\[%
\begin{array}
[c]{cc}%
{\includegraphics[
height=2.7223in,
width=3.4811in
]%
{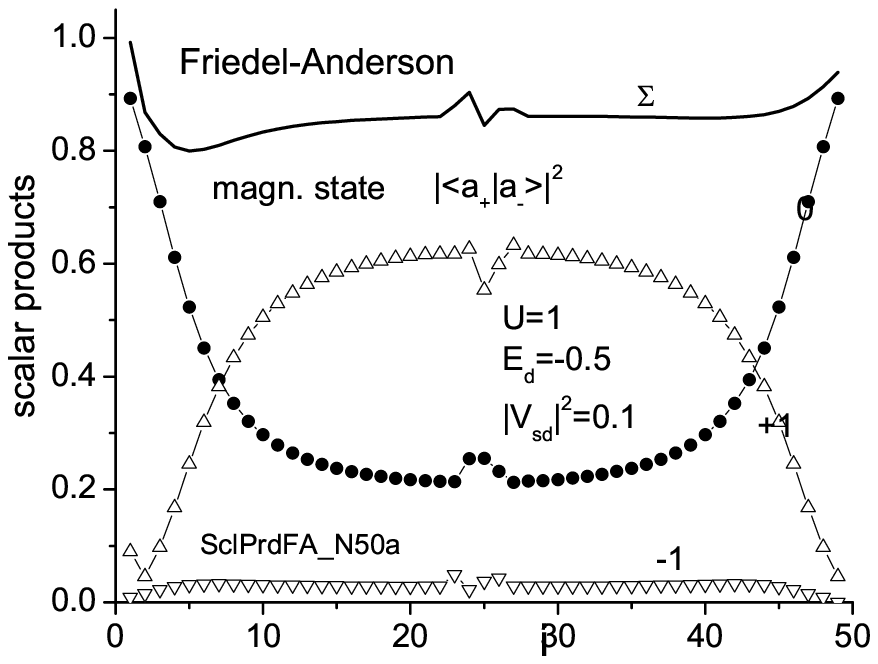}%
}%
&
\end{array}%
\begin{tabular}
[c]{l}%
Fig.4: The square of the diagonal\\
matrix-elements $\left\vert \left\langle a_{+,i}^{\dag}\Phi_{0}|a_{-,i}^{\dag
}\Phi_{0}\right\rangle \right\vert ^{2}$\\
as well as the next-to-diagonal\\
matrix-elements $\left\vert \left\langle a_{+,i}^{\dag}\Phi_{0}|a_{-,i\pm
1}^{\dag}\Phi_{0}\right\rangle \right\vert ^{2}$\\
are plotted as a function of $i$ for the\\
enforced magnetic state.
\end{tabular}
\ \
\]

\[
\]

As a consequence the two bases $\left\{  a_{+,i}^{\dag}\right\}  $ and
$\left\{  a_{-,i}^{\dag}\right\}  $ are quite similar. In Fig.4 the
single-particle scalar products $\left\vert \left\langle a_{+,i}^{\dag}%
\Phi_{0}|a_{-,j}^{\dag}\Phi_{0}\right\rangle \right\vert ^{2}$ for the
enforced magnetic state are plotted as a function of $i$ (full circles). In
addition $\left\vert \left\langle a_{+,i}^{\dag}\Phi_{0}|a_{-,i\pm1}^{\dag
}\Phi_{0}\right\rangle \right\vert ^{2}$ are plotted as empty up and down
triangles. The full curve (without symbols) gives the sum of the three
contributions. One recognizes that an arbitrary state $a_{+,i}^{\dag}$ (for
$i>0$) can be constructed to 80\% out of the states $a_{-,i}^{\dag}%
,a_{-,i+1}^{\dag}$ and $a_{-,i-1}^{\dag}$. On the other hand $a_{+,i}^{\dag}$
and $a_{-,i}^{\dag}$ overlap to only 30\% for small energies (in the center of
the horizontal axis).%
\[%
\begin{array}
[c]{cc}%
{\includegraphics[
height=2.9439in,
width=3.7642in
]%
{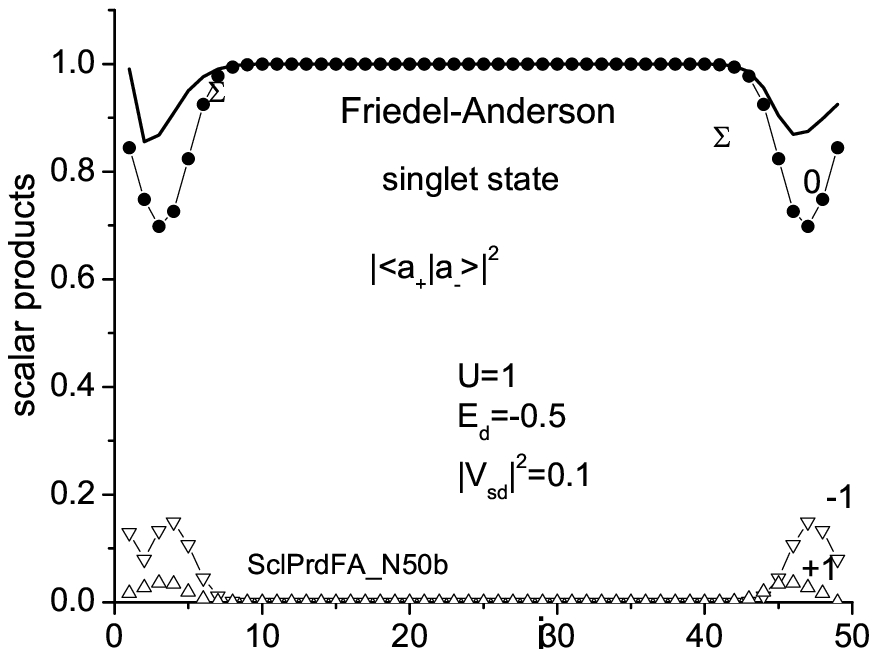}%
}%
&
\end{array}%
\begin{tabular}
[c]{l}%
Fig.5: The square of the diagonal\\
matrix-elements $\left\vert \left\langle a_{+,i}^{\dag}\Phi_{0}|a_{-,i}^{\dag
}\Phi_{0}\right\rangle \right\vert ^{2}$\\
as well as the next-to-diagonal\\
matrix-elements $\left\vert \left\langle a_{+,i}^{\dag}\Phi_{0}|a_{-,i\pm
1}^{\dag}\Phi_{0}\right\rangle \right\vert ^{2}$\\
are plotted as a function of $i$ for the\\
singlet state.
\end{tabular}
\ \ \
\]

\[
\]

This is very different for the singlet state. In Fig.5 the single-electron
scalar products $\left\vert \left\langle a_{+,i}^{\dag}\Phi_{0}|a_{-,i}^{\dag
}\Phi_{0}\right\rangle \right\vert ^{2}$ as well as $\left\vert \left\langle
a_{+,i}^{\dag}\Phi_{0}|a_{-,i\pm1}^{\dag}\Phi_{0}\right\rangle \right\vert
^{2}$ are plotted as a function of $i$ for the singlet state. One recognizes
that over a large energy range the states $a_{+,i}^{\dag}$ and $a_{-,i}^{\dag
}$ are 99\% or more identical. Only for (absolute) larger energies on the left
and right side is the overlap reduced to about 70\%.

The reason for this different behavior is rather transparent. The energy
expectation value of the enforced magnetic state does not depend on the MESP.
The s-d transitions happen only within the same spin orientation and therefore
between the same bases. As an example, one has the transition
\[
a_{0-\downarrow}^{\dag}d_{\uparrow}^{\dag}\prod_{i=1}^{n-1}a_{i-\downarrow
}^{\dag}\prod_{i=1}^{n-1}a_{i+\uparrow}^{\dag}\Phi_{0}<=>d_{\downarrow}^{\dag
}d_{\uparrow}^{\dag}\prod_{i=1}^{n-1}a_{i-\downarrow}^{\dag}\prod_{i=1}%
^{n-1}a_{i+\uparrow}^{\dag}\Phi_{0}%
\]
Here the matrix element is just
\begin{align*}
&  \left\langle a_{0-\downarrow}^{\dag}d_{\uparrow}^{\dag}\prod_{i=1}%
^{n-1}a_{i-\downarrow}^{\dag}\prod_{i=1}^{n-1}a_{i+\uparrow}^{\dag}\Phi
_{0}\left\vert V_{sd}^{-}(0)a_{0,\downarrow}^{\dag}d_{\downarrow}]\right\vert
d_{\downarrow}^{\dag}d_{\uparrow}^{\dag}\prod_{i=1}^{n-1}a_{i-\downarrow
}^{\dag}\prod_{i=1}^{n-1}a_{i+\uparrow}^{\dag}\Phi_{0}\right\rangle \\
&  =\left\langle a_{0-\downarrow}^{\dag}\Phi_{0}\left\vert V_{sd}%
^{-}(0)a_{0,\downarrow}^{\dag}d_{\downarrow}]\right\vert d_{\downarrow}^{\dag
}\Phi_{0}\right\rangle =V_{sd}^{-}\left(  0\right)
\end{align*}
With $a_{0-}^{\dag}=%
{\textstyle\sum_{\nu=1}^{N}}
\alpha_{0-}^{\nu}c_{\nu}^{\dag}$ the value of $V_{sd}^{-}\left(  0\right)  $
is given by%
\[
V_{sd}^{-}\left(  0\right)  =%
{\textstyle\sum_{\nu=1}^{N}}
\alpha_{0-}^{\nu}V_{sd}\left(  \nu\right)
\]
There are no processes that involve the MESP.

On the other hand in the singlet state one has transitions from
\[
a_{0-\downarrow}^{\dag}a_{0+\uparrow}^{\dag}\prod_{i=1}^{n-1}a_{i-\downarrow
}^{\dag}\prod_{i=1}^{n-1}a_{i+\uparrow}^{\dag}\Phi_{0}<=>d_{\uparrow}^{\dag
}a_{0+\downarrow}^{\dag}\prod_{i=1}^{n-1}a_{i-\uparrow}^{\dag}\prod
_{i=1}^{n-1}a_{i+\downarrow}^{\dag}\Phi_{0}%
\]
Such a transition is proportional to the square of the MESP. To be able to
harvest energy from these processes the states $a_{i+}^{\dag}$ and
$a_{i-}^{\dag}$ are, for small energies, aligned parallel to each other.

In Fig.6 the corresponding single-particle scalar products $\left\vert
\left\langle a_{+,i}^{\dag}\Phi_{0}|a_{-,j}^{\dag}\Phi_{0}\right\rangle
\right\vert ^{2}$ for the triplet state are plotted as a function of $i$. With
exception of the few (2 - 4) points in the center the results is very close to
the result of the enforced magnetic state. The two (magnetic) components of
the triplet state are essentially decoupled because the coupling is
proportional to $\left[  M^{(N/2}\right]  ^{2}$ which is of the order of
$10^{-7}$. Therefore the triplet state is essentially equal to the sum of the
enforced magnet state plus its time- (spin-) reversed partner (for
sufficiently large $N$).%
\[%
\begin{array}
[c]{cc}%
{\includegraphics[
height=3.1474in,
width=4.0174in
]%
{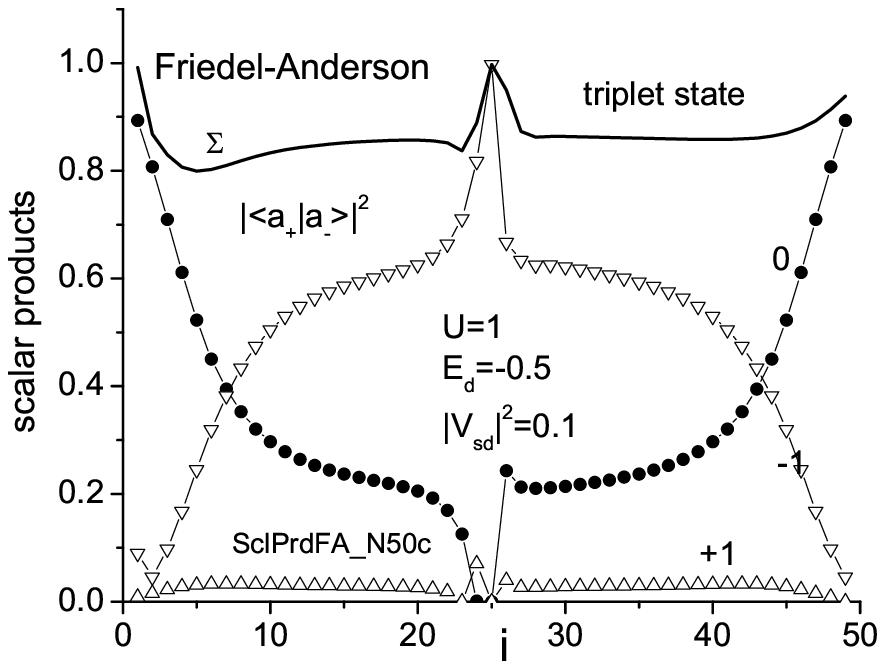}%
}%
&
\end{array}%
\begin{tabular}
[c]{l}%
Fig.6: The square of the diagonal\\
matrix-elements $\left\vert \left\langle a_{+,i}^{\dag}\Phi_{0}|a_{-,i}^{\dag
}\Phi_{0}\right\rangle \right\vert ^{2}$\\
as well as the next-to diagonal\\
matrix-elements $\left\vert \left\langle a_{+,i}^{\dag}\Phi_{0}|a_{-,i\pm
1}^{\dag}\Phi_{0}\right\rangle \right\vert ^{2}$\\
are plotted as a function of $i$ for the\\
triplet state.
\end{tabular}
\
\]

The results of this paper are two-fold. One has to distinguish whether (the
expectation value of) the energy in the ground state depends on the MESP. If
$E_{0}$ is independent of the MESP and one uses the Wilson states as the
original basis then the MESP $M^{N/2}$ decreases exponentially with increasing
$N$. This is due to the fact that the energy of the states closest to the
Fermi level decreases exponentially as well. Just by moving these states
closer decreases $M^{N/2}$. On the other hand if the total number of states in
a given energy interval is increased without reducing the energy of the states
closest to the Fermi energy then the MESP $M^{N/2}$ is barely affected.

If the energy in the ground state depends on the MESP (as for the singlet
state) then this results in a freeze of the MESP. This achieved by forcing the
new basis states $a_{i,+}^{\dag}$ and $a_{i,-}^{\dag}$ to be parallel within a
certain energy range of the Fermi level.
\[
\]

\section{Appendix}

\appendix{}

\section{Wilson states}

Wilson \cite{W18} in his Kondo paper considered an s-band ranging from $-1$ to
1 with a constant density of states. In the next step Wilson replaced the
energy continuum of s-states by a discrete set of cells. First the negative
energy band is subdivided on a logarithmic scale. The discrete energy values
are $-1,-1/2,-1/2^{2}$,$-2^{-\nu},..-2^{-\left(  N/2-1\right)  },0$. These
discrete $\xi_{\nu}=-2^{-\nu}$ points are used to define a sequence of energy
cells: the cell $\mathfrak{C}_{\nu}$ (for $\nu$%
$<$%
N/2) includes all states within $\left(  \xi_{\nu-1}:\xi_{\nu}\right)
=\left(  -1/2^{\nu-1}:-1/2^{\nu}\right)  $. A new (Wilson) state $c_{\nu
}^{\dag}$ is a superposition of all states within an energy cell $\left(
\xi_{\nu-1}:\xi_{\nu}\right)  $ and has an (averaged) energy $\left(  \xi
_{\nu-1}+\xi_{\nu}\right)  /2=\allowbreak\left(  -\dfrac{3}{2}\right)
\dfrac{1}{2^{\nu}}$. This yields a spectrum $\varepsilon_{\nu}$: $-\frac{3}%
{4},-\frac{3}{8},-\frac{3}{16},$ $..,-\frac{3}{2^{N/2}},-\frac{1}{2^{N/2}}$.
This spectrum is extended symmetrically to positive energies (for $\nu>N/2$).
The essential advantage of the Wilson basis is that it has an arbitrarily fine
energy spacing at the Fermi energy. For a given $N$ the two smallest energy
cells extend from $\pm2^{-\left(  N/2-1\right)  }$ to $0,$ and the (absolute)
smallest energy levels are $\pm2^{-N/2}$.%
\[
\]

\end{document}